# Superconductivity in the iron selenide $K_xFe_2Se_2$ ($0 \leq x \leq 1$)

**Jiangang Guo [†], Shifeng Jin [†], Gang Wang [†], Shunchong Wang [†], Kaixing Zhu[†], Tingting Zhou[†], Meng He[‡], Xiaolong Chen[†,*]**

[†] Research & Development Center for Functional Crystals, Beijing National Laboratory for Condensed Matter Physics, Institute of Physics, Chinese Academy of Sciences, P.O. Box 603, Beijing 100190, China

[‡] National Centre for Nanoscience and Technology, Beijing 100190, China

**Abstract**

We report the superconductivity at above 30 K in a new FeSe-layer compound $K_{0.8}Fe_2Se_2$ (nominal composition) achieved by metal K intercalating in between FeSe layers. It is isostructural to $BaFe_2As_2$ and possesses the highest $T_c$ for FeSe-layer materials so far under ambient pressure. Hall effect indicates the carriers are dominated by electron in this superconductor. We confirm that the observed superconductivity at above 30 K is due to this new FeSe-based 122 phase. Our results demonstrate that FeSe-layer materials are really remarkable superconductors via structure and carrier modulation.

The discovery of superconductivity in the quaternary ZrCuSiAs-type (referred to as 1111-type) Fe-based oxypnictide LaFeAsO (F-doped) with critical temperature ($T_c$) 26 K has inspired worldwide interests on high temperature superconductors once more.[1] After that, a series of new superconducting materials based on FeAs-layer were found, such as $AFe_2As_2$ (A=K, Sr, Ba),[2-4] LiFeAs,[5] and $Sr_2VO_3FeAs$,[6] which contain As-$Fe_2$-As layers composed of edge-sharing $FeAs_4$ tetrahedra. Up to now, the highest reported 55 K for iron-based superconductivity was achieved in oxygen-deficient

SmFeAsO prepared by high pressure synthesis.[7] The similarities between the iron-based and copper-based superconductors, including high $T_c$ values, and proximity of a magnetically ordered state, suggest that the superconductivity in the Fe-based materials is unconventional and non-BCS-like.[8-10]

Recently, superconductivity has been reported at 8 K in the structurally related material anti-PbO-type FeSe (referred to as 11-type).[11] Compared to the iron pnictides, FeSe has a substantially simplified structure stacking of only FeSe layers and no intercalating cations. It has no static magnetic ordering under pressures up to 38 GPa, differing from other iron-based superconductors.[12] More recently, the angle-resolved photoemission spectroscopy (ARPES) study demonstrated that the normal state of FeSe$_{0.42}$Te$_{0.58}$ is a strongly correlated metal and significantly different from the 1111 and 122 iron-pnictides families in electronic properties.[13] Moreover, by tellurium doping or the exertion of high pressure, the superconducting transition temperature of FeSe can even increase up to 15.2 K and 37 K, respectively.[14, 15] These results establish that the FeSe-layer materials, in addition to serving as test objects for study of physics of pnictides, could themselves prove to be remarkable superconductors.[16] In this study, we report the superconductivity at above 30 K in a new FeSe-layer compound K$_{0.8}$Fe$_2$Se$_2$ (nominal composition) achieved by alkali metal intercalating in between FeSe layers. It is the reported highest value of $T_c$ for FeSe-layer materials so far under ambient pressure. Our results demonstrate that FeSe-layer materials are really remarkable superconductors with high $T_c$ via structure and carrier modulation.

A series of polycrystalline samples were synthesized using a two-step solid state reaction method. First, FeSe powders were prepared with high purity powder of selenium (Alfa, 99.99%) and iron (Alfa, 99.9+%) by a similar method to that described in Ref. 11. Then, FeSe and K (Sinopharm Chemical, 97%) mixed with appropriate stoichiometry were heated in alumina crucibles, sealed in quartz tubes partially backfilled with ultra-high-purity argon. The samples were heated to 973-1023 K, kept at the temperatures for 30 h and cooled naturally to room temperature by switching off the furnace. To get the samples with better crystallinity, we tried to grow single crystal by flux method. The precursors of the same stoichiometry were put into alumina crucibles and sealed in silica ampoules under an atmosphere of 0.2 bar argon. The samples were heated to 1300 K, cooled down to 1000 K at the rate of 4 K/hr and finally furnace-cooled to room temperature. Well-formed plates with shiny surfaces having dimensions up to 3 mm × 2 mm × 0.5 mm were obtained. The as-prepared samples were characterized by powder X-ray diffraction (PXRD) using a panalytical X'pert diffractometer with Cu $K_a$ radiation. Rietveld refinements of the data were performed with the FULLPROF package.[17] The electrical resistance and Hall coefficient were measured through the standard four-wire method and five-wire technique on the physical property measurement system (PPMS, Quantum Design), respectively. The dc magnetic properties were characterized using a vibrating sample magnetometer (VSM, Quantum Design). It should be noted that the electrical and magnetic properties were measured based on crystals.

Figure 1 shows the diffraction pattern of polycrystalline sample collected at room temperature. All the reflections could be indexed with lattice parameters $a = b =$ 3.9136 (1) Å, $c =$ 14.0367 (7) Å and $V =$ 214.991(13) Å$^3$. The systematic absence of *hkl* suggests that the space group probably be *I*4/*mmm* (No. 139). The intercalation of K in between FeSe layers is clearly proven by the remarkable increase in lattice parameter *c* compared with that of FeSe.[11] Rietveld refinements were then performed by adopting a structural analogue of KCo$_2$Se$_2$ as an initial model.[18] The refinements smoothly converged to $R_p =$ 3.26 %, $R_{wp}=$ 5.15%, and $R_{exp}=$ 2.22%, respectively. A summary of the crystallographic data is compiled in TableⅠ. The crystal structure of KFe$_2$Se$_2$, as shown schematically in inset of Fig. 1, is composed of antifluorite-type Se-Fe$_2$-Se layers of edge-sharing FeSe$_4$ tetrahedra separated by K cations, which is identical to the well-known ThCr$_2$Si$_2$-structure. The (Fe$_2$Se$_2$)$^{\delta-}$ layers serve as the "conducting layer" and K$^+$ ions provide charge carriers, quite alike to other 122 structures. Compared with the structural data for FeSe,[19] the intralayer Fe-Fe distance and Fe-Se bond distance increases by 3.7%, and 2.15%, respectively. The interlayer distance of two neighboring Fe-Fe square plane, 7.0184(5) Å, is the largest discovered so far, a consequence of K intercalated into FeSe layers. The increase in *a*-axis is significantly smaller than that of *c*-axis, which leads to the reduced dimensionality in KFe$_2$Se$_2$. As one of the most important structural features, the refinement indicates that the Fe-Se-Fe angle is 110.926(4)°, which is closer to the ideal angle of high symmetric tetrahedra compared to that of FeSe. The X-ray diffraction pattern of crystal (as shown in Fig. 2) is dominated by the 00*l* (*l* = 2n) reflections, suggesting that the cleave surface of the plate-like crystal is approximately perpendicular to the crystallographic *c*-axis. The elemental composition was checked with several crystals from the same boule by Inductively Coupled Plasma - atomic emission spectrometer (ICP-AES). The chemical analyses show that the average atomic ratios of K: Fe: Se is 0.39:0.85:1, a little deficiency of potassium and iron.

Figure 3 displays the temperature dependence of in-plane electrical resistance ($R_{ab}$) of $K_{0.8}Fe_2Se_2$ crystal. The resistance shows a broad peak centered at 105 K and exhibits semiconducting characteristics in the half of higher temperature range. As the temperature is further decreased, the resistance displays a metallic behavior and drops abruptly at about 30 K, which clearly indicates superconductivity. Lower inset of Fig. 3 shows details of the transition. The onset transition temperature by extrapolating is at 30.1 K, and zero resistance is achieved at 27.2 K. By 90/10 criterion, we find the midpoint of the resistive transition where the resistance drops to 50% of that of the onset at 28.3 K. and a transition width of ~1.3 K, suggesting the high quality and the homogeneous nature of the crystal. The Hall coefficient $R_H$ (higher inset of Fig. 3) is negative over the whole temperature range from 70 to 250 K, indicating the conduction carriers are dominated by electrons. For a single-band model, the carrier concentration is roughly estimated to be $n \approx 1.76 \times 10^{21}$ cm$^{-3}$ from the Hall coefficient by $n=1/(R_He)$ at 70 K, which is a little less than that of $Na_{1-\delta}FeAs$.[20] Above 105 K, $R_H$ is nearly temperature independent. $R_H$ drops dramatically below 105 K, which is the exactly same temperature as the peak position observed in the resistive measurement. As FeAs-based superconductors, the strong temperature dependence below 105 K is qualitatively anticipated by the gradual forming of gap in the Fermi surface.[6, 20]

The magnetization of $K_{0.8}Fe_2Se_2$ crystal as a function of temperature was measured at a magnetic field of $H$ = 50 Oe, as shown in Fig. 4. In the temperature range from room temperature to the onset transition temperature, the zero field cooled (ZFC) and field cooled (FC) magnetization curves are essentially flat and temperature independent, indicating that the sample is a Pauli paramagnet. No magnetic anomaly was detected around 105 K. As shown in the left inset of Fig. 4, a clear diamagnetic response appears at 31 K, which occurs at almost the same temperature for the electrical resistance. The existing hysteresis between ZFC and FC curves indicates that the material is a typical type-II superconductor, which is the same as other iron-based superconductors and the copper-based superconductors. The superconducting volume fraction estimated from the ZFC magnetization at 10 K is ~60% (calculated as from the value of perfect diamagnetism). The right inset of Fig. 4 shows the *M-H* curve at 5 K. It also indicates a typical profile for type-II superconductor. From the *M-H* curve, the lower critical magnetic field ($H_{c1}$) is around 0.2 T and the estimated upper critical magnetic field ($H_{c2}$) is higher than 9 T. These observations collectively confirm the bulk superconductivity nature and the existence of strong-coupling in $K_{0.8}Fe_2Se_2$ crystal.

To investigate the origin of superconductivity, we first take account of impurity, such as FeSe, which is critical in confirming the source of superconductivity in the obtained samples. Here, the most convincing argument against the effect of FeSe is the XRD patterns of $K_xFe_2Se_2$ (0≤x≤1.0) polycrystalline samples. The diffraction patterns of $K_xFe_2Se_2$ (0≤x≤1.0) with different K contents are shown in Fig. 5a. With increasing K contents, the FeSe (101) reflection gradually weakens and the $KFe_2Se_2$ (103) reflection gradually enhances. When x is above 0.8, the FeSe (101) reflection finally disappears, while when x below this value but above zero, the samples contain both $KFe_2Se_2$ and FeSe. For x = 0.4, the magnetization [Fig. 5b] clearly exhibits two magnetic singularities at ~8 K and ~31 K, respectively. The sharp drop occurring at ~8 K is attributed to the superconducting transition of FeSe, while the magnetic anomaly at ~31 K is correlated with $KFe_2Se_2$. The relatively large positive background of magnetization can be possibly ascribed to the existence of Fe impurity. As x increases up to 0.8, the singularity at ~8 K totally disappears and only the magnetic anomaly at ~31 K left. It rules out the possibility that the superconductivity originates from FeSe. Thus, we conclude that the superconductivity at above 30 K can be only contributed to $K_{0.8}Fe_2Se_2$.

The higher $T_c$ in this new superconductor may be related to its structural feature and electron doping. Structural refinement reveals that the Se-Fe-Se bond angle in $KFe_2Se_2$ is significantly close to the ideal $FeSe_4$ tetrahedral shape and the interlayer distance is dramatically large compared with those of FeSe under ambient pressure as well as under high pressure. The Se-Fe-Se bond angle of FeSe is 112.096º at ambient pressure, and increases smoothly with increasing pressure, implying the bigger distortion emerges under high pressure.[15] Simultaneously, the separation of SeFeSe building blocks in FeSe is ~2.55 Å, which is smaller than 4.1013(3) Å of $KFe_2Se_2$. No matter under ambient or high pressure, these structural features of FeSe contrast sharply with the regular tetrahedral geometry and enhanced interlayer separation existing in KFe$_2$Se$_2$. However, the enhanced superconductivity in K$_{0.8}$Fe$_2$Se$_2$ seems to obey the empirical rule among FeAs-based superconductors, that is, the materials with perfect FeAs$_4$ tetrahedra always approach to the highest T$_c$.[21, 22] If we assume that is also applicable to FeSe-layer system, the less structural distortion from FeSe$_4$ tetrahedra in KFe$_2$Se$_2$ might induce the enhancement of density of states near the Fermi energy and optimal effective exchange couplings of Fe-Se and nearest-neighbor Fe-Fe. Thus, from the viewpoint of crystal structure, the optimal electronic structure existing in more regular FeSe$_4$ tetrahedra shape is likely to enhance superconductivity in FeSe-layer system.

In conclusion, we have discovered a new FeSe-based layered superconductor K$_{0.8}$Fe$_2$Se$_2$ with T$_c$ above 30 K, which is the highest value among FeSe-layer compounds at ambient pressure so far. Structural studies reveal the FeSe$_4$ tetrahedra in the FeSe sheets are closer to the ideal tetrahedra, though its Fe-Fe, Fe-Se bond lengths are enlarged after the K intercalation. On the other hand, in this superconductor, the carrier is dominated by electron, and its density is ~$10^{21}$/cm$^3$. This value is a little bit lower in comparison of that in FeSe. This may imply the spin fluctuations be enhanced in KFe$_2$Se$_2$ by intercalating K. We found the transport properties are sensitive to the compositions and synthesizing conditions. Efforts in optimization are underway to unravel the intrinsic superconducting properties.

This work was partly supported by the National Natural Science Foundation of China under Grants No. 90922037, 50872144 and 50972162 and the International

Centre for Diffraction Data (ICDD, USA).

Table and Figure Captions:

TABLE I. Crystallographic data of $KFe_2Se_2$.

FIG. 1. (Color online) Powder x-ray diffraction and Ritveld refinement profile of $KFe_2Se_2$ at 297 K. The inset shows the schematic crystal structure of $KFe_2Se_2$ ($ThCr_2Si_2$-type).

FIG. 2. (Color online) The x-ray diffraction pattern of $K_{0.8}Fe_2Se_2$ crystal indicates that the (00$l$) ($l$ = 2n) reflections dominate the pattern. The asterisk shows an unknown reflection. Inset shows the photography of the $K_{0.8}Fe_2Se_2$ crystal (length scale 1 mm).

FIG. 3. The temperature dependence of electrical resistance for the $K_{0.8}Fe_2Se_2$ crystal sample. The lower inset shows the details of superconducting transition from 10 K to 40 K. The upper inset shows temperature dependence of normal state Hall coefficient for crystal sample.

FIG. 4. (Color online) The magnetization of $K_{0.8}Fe_2Se_2$ crystal as a function of temperature with the $H$ parallel to $c$ axis. The left inset shows the expanded view of the temperature dependence of the magnetization near the onset of superconducting transition. The right left inset shows the magnetization versus $H$ at 5 K.

FIG. 5. (Color online) (a) The intercalated concentration dependence of PXRD peaks of the FeSe (101) and the $K_xFe_2Se_2$ (103) reflections collected at room temperature. The inset shows the powder x-ray diffraction patterns for $K_xFe_2Se_2$ (0≤x≤1.0) polycrystalline samples at room temperature. Dots mark the reflections of FeSe. (b) Temperature dependent magnetization for representative member $K_{0.4}Fe_2Se_2$ polycrystalline sample (arrow shows the transition point).

| | |
|---|---|
| Formula | KFe$_2$Se$_2$ |
| Temperature (K) | 297 |
| Space group | *I4/mmm* |
| Fw | 212.03 |
| *a* (Å) | 3.9136(1) |
| *c* (Å) | 14.0367(7) |
| V (Å$^3$) | 214.991(3) |
| Z | 2 |
| $R_p$ | 3.26% |
| $R_{wp}$ | 5.15% |
| $R_{exp}$ | 2.22% |
| $\chi^2$ | 5.38 |
| Atomic parameters: | |
| K | 2a (0, 0, 0) |
| Fe | 4d (0, 0.5, 0.25) |
| Se | 4e (0, 0, z) |
| | z=0.3539(2) |
| Bond length (Å): | |
| K-Se | 3.4443(4) × 8 |
| Fe-Se | 2.4406(4) × 4 |
| Fe-Fe | 2.7673(5) × 4 |
| Bond angles (deg): | |
| | 110.926(4) × 4 |
| | 106.600(4) × 2 |

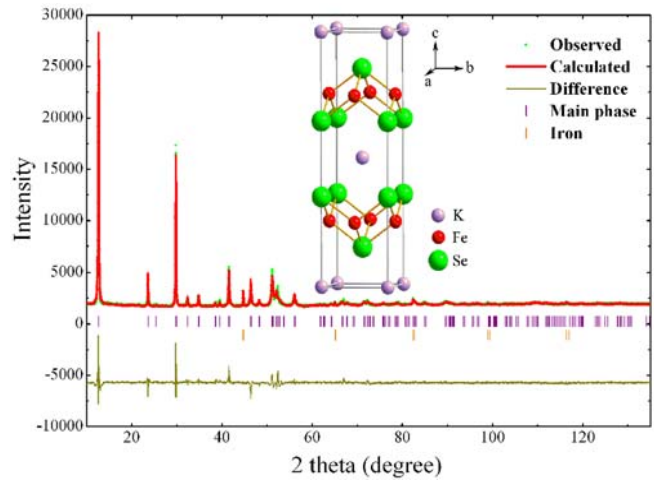

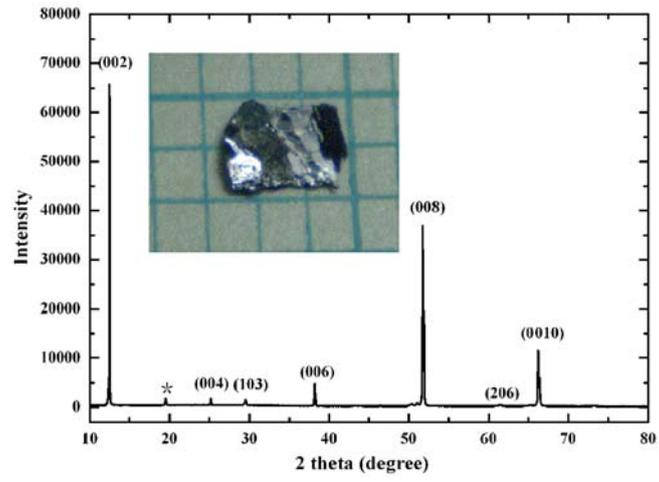

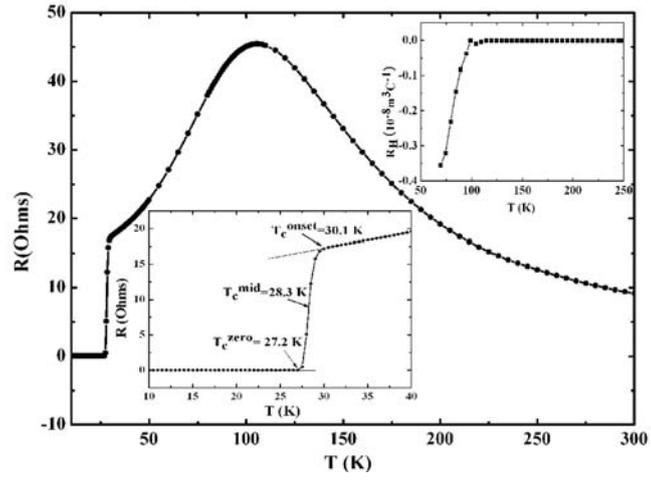

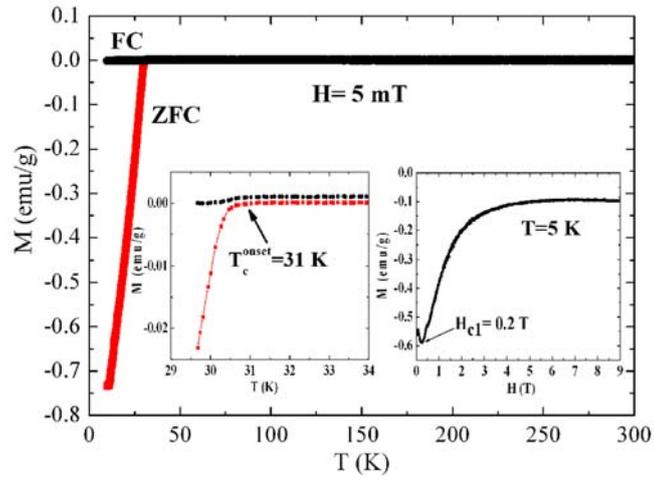

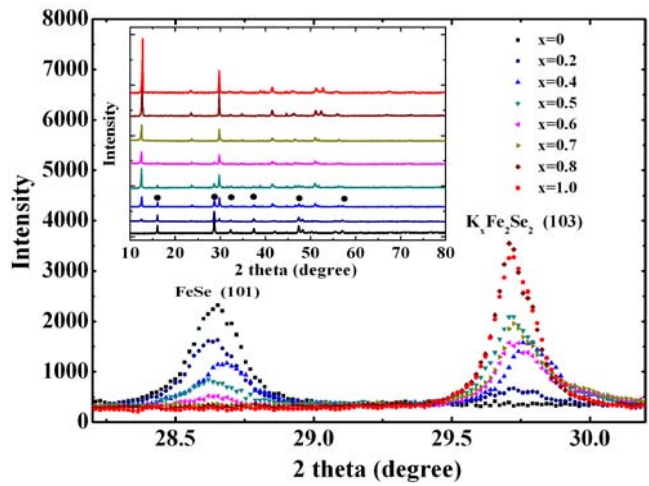

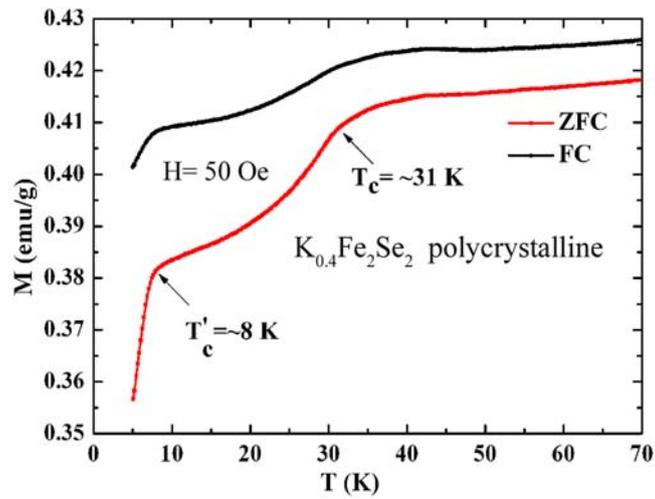